\documentclass[aps,prb,twocolumn,superscriptaddress]{revtex4-1}
\setcitestyle{authoryear,round}
\usepackage{graphicx}
\usepackage{natbib}
\usepackage{bm}
\usepackage{amsmath}
\usepackage{amssymb}
\usepackage{hyperref}
\usepackage[version=4]{mhchem}
\usepackage{color}
\usepackage{upgreek}

\begin{document}

\title{Machine Learning in Stellar Astronomy: Progress up to 2024}

\author{Guangping Li}
\affiliation{Laboratory for Relativistic Astrophysics, Department of Physics, Guangxi University, Nanning 530004, China}

\author{Zujia Lu}
\affiliation{Laboratory for Relativistic Astrophysics, Department of Physics, Guangxi University, Nanning 530004, China}

\author{Junzhi Wang}
\affiliation{Laboratory for Relativistic Astrophysics, Department of Physics, Guangxi University, Nanning 530004, China}

\author{Zhao Wang}
\email{zw@gxu.edu.cn}
\affiliation{Laboratory for Relativistic Astrophysics, Department of Physics, Guangxi University, Nanning 530004, China}

\begin{abstract}

Machine learning (ML) has become a key tool in astronomy, driving advancements in the analysis and interpretation of complex datasets from observations. This article reviews the application of ML techniques in the identification and classification of stellar objects, alongside the inference of their key astrophysical properties. We highlight the role of both supervised and unsupervised ML algorithms, particularly deep learning models, in classifying stars and enhancing our understanding of essential stellar parameters, such as mass, age, and chemical composition. We discuss ML applications in the study of various stellar objects, including binaries, supernovae, dwarfs, young stellar objects, variables, metal-poor, and chemically peculiar stars. Additionally, we examine the role of ML in investigating star-related interstellar medium objects, such as protoplanetary disks, planetary nebulae, cold neutral medium, feedback bubbles, and molecular clouds.

\end{abstract}


\maketitle

\section{Introduction}
\label{sect:intro}
Stellar astronomy, which involves the study of star characteristics, dynamics, and lifecycles, has undergone substantial transformation in recent decades due to the development of advanced instruments, as exemplified by the deployment of cutting-edge observatories. These instruments have enabled unprecedented precision and depth in astronomical observations. As a result, modern astronomy is now dominated by large-scale, high-resolution datasets, which have brought both exciting opportunities and significant challenges to the field. The massive influx of high-quality observational data has propelled stellar astronomy into the ``big data'' era, where the ability to manage, analyze, and interpret immense volumes of information is critical \citep{Sharma2017}. Traditional methods, which often rely on manual inspection and rule-based techniques, are increasingly insufficient to handle the sheer scale and complexity of contemporary astronomical datasets. These limitations are most apparent in the difficulties associated with managing the enormous volumes of data generated by current surveys and the inability to easily identify subtle, multidimensional correlations between astrophysical parameters.

In response to these challenges, machine learning (ML), a subset of artificial intelligence, has emerged as a powerful tool, offering advanced algorithms capable of efficiently processing large, multidimensional datasets \citep{Meher2021,Smith2023}. ML has proven particularly valuable in automating tasks such as the classification of stellar objects, inference of astrophysical parameters, and pattern recognition in noisy data. By leveraging complex algorithms, including deep learning, ML techniques are able to outperform traditional methods in both accuracy and speed, enabling the extraction of meaningful insights from astronomical data that would otherwise be challenging to uncover \citep{Rodriguez2022}.

\begin{figure*}[htb]
    \centering
    \includegraphics[width=0.95\textwidth]{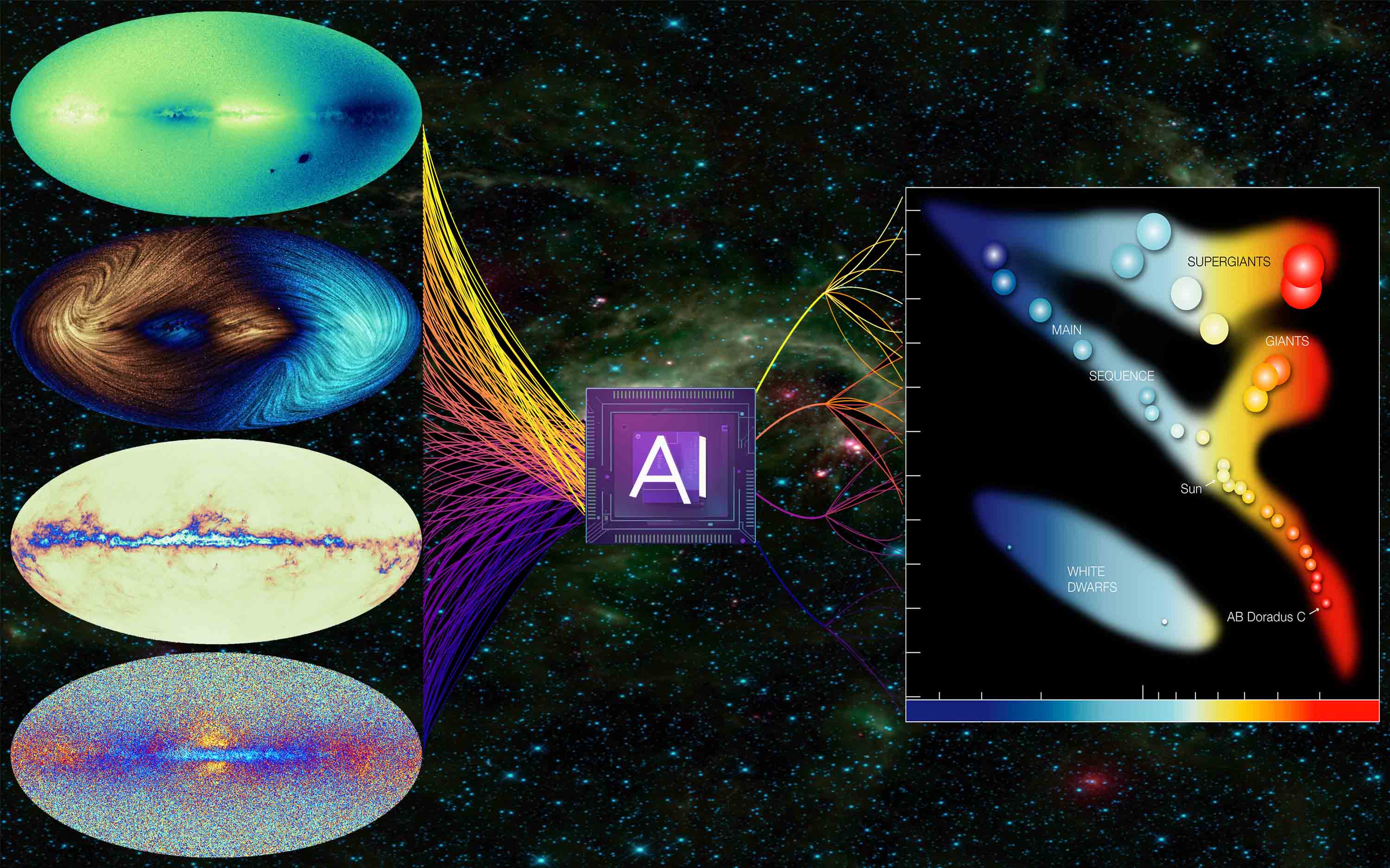}
    \caption{Schematics illustrating the role of AI in advancing our understanding of star evolution. Credit for parts of the image: \textcopyright{ESA} and NASA/JPL-Caltech/UCLA.}
    \label{F0}
\end{figure*}

Historically, the classification and identification of stellar objects required extensive classification efforts based on spectral or photometric properties \citep{Hernandez2004}, which was time consuming and prone to human error. Modern ML algorithms, particularly neural networks, have shown remarkable success in automating this process, allowing for the rapid and precise classification of different types of stars, even when dealing with data with low signal-to-noise ratios \citep{Sen2022}. This capability is becoming increasingly important as sky surveys grow in scale, with projects such as the Sloan Digital Sky Survey(SDSS) and the European Space Agency's Gaia mission generating catalogs containing billions of stars. ML algorithms have helped classifying stars at different stages of evolution, facilitating a more precise mapping of stars on the Hertzsprung-Russell diagram, as shown in Figure \ref{F0}. By leveraging large datasets such as those from Gaia, ML techniques provide valuable insights into stellar evolution, deepening our understanding of how stars form, evolve, and ultimately reach the end of their life cycles.

Beyond object classification, ML techniques are playing a critical role in the extraction of key astrophysical parameters from observational data. In many cases, the relationships between observable properties (such as spectral energy distributions) and underlying stellar characteristics (such as mass, temperature, and age) are highly nonlinear and involve multiple interacting variables \citep{woosley2002}. ML models have been employed to infer these parameters with high precision. Supervised learning techniques, for example, have been used to derive effective temperatures, surface gravities, and chemical abundances from spectroscopic and photometric data \citep{Yu2021}. These techniques are leading to new insights into stellar populations, their evolution, and their roles in the broader context of galactic evolution.

ML is helping to address fundamental questions related to the lifecycle of stars, their intrinsic properties, and their influence on the interstellar medium (ISM). These inquiries, central to the discipline, often require the analysis of highly complex, multidimensional datasets, where traditional methods face significant limitations. For instance, understanding stellar evolution involves deciphering non-linear relationships between observable features and underlying stellar parameters, a task where supervised ML algorithms, such as those employed by \cite{Jones2000,Bellinger2016}, have demonstrated considerable utility. Similarly, the study of stellar populations, as exemplified by \citep{Garraffo2021}, highlights the capacity of deep neural networks to infer stellar masses and ages with a level of precision that was previously unattainable by manual or rule-based approaches.

The role of ML extends to probing the interplay between stars and their surrounding interstellar environments. For example, the identification of massive stars, a task successfully addressed through photometric classifiers by \cite{Maravelias2022}, provides crucial insights into feedback processes that shape star formation and galactic dynamics 
. Additionally, the extraction of spectral properties through neural networks, as explored by  \cite{Skoda2020}, facilitates the detection of stars with emission-line features, which are often associated with active or young stellar environments 
. These advancements underscore the ability of ML techniques to enhance our understanding of the ISM by enabling the efficient characterization of stellar sources that influence their surroundings.

However, despite the considerable promise of ML applications, challenges remain. One of the primary concerns is the dependence of the ML models on the quality of the training data. Poorly curated or biased datasets can result in inaccurate or misleading results, especially when models are applied beyond the scope of their training sets. Furthermore, ML models, particularly deep learning models, are often criticized for their ``black box'' nature. The lack of interpretability in these models makes it difficult to understand the underlying physics driving their predictions, which can hinder scientific progress. As a result, ongoing efforts aim to develop more interpretable ML models in the context of astrophysical applications.

\section {Scope and Outline}

To provide a detailed overview of the progress made in this area, this paper offers a comprehensive summary of ML applications in stellar astronomy. Covering the identification and classification of stellar objects as well as the inference of astrophysical parameters, this article is organized as follows. Section 3 introduces the fundamentals of ML, while Section 4 provides an overview of commonly used observational data sources that lay the foundation for ML applications. Section 5 summarizes ML studies on general star identification and classification, and Section 6 explores research focused on inferring astrophysical parameters using ML. Section 7 surveys ML studies targeting specific types of stellar objects, including binaries, supernovae, dwarf stars, young stellar objects, variables, metal-poor stars, chemically peculiar stars, and relevant ISM objects. Finally, Section 8 discusses challenges and future directions and Section 9 offers concluding remarks.

Although this article aims to provide a broad overview, some discussions may remain general or superficial because of the vast amount of information and the diverse research directions in stellar astronomy. It also seems premature to fully appreciate the key roles of ML in addressing critical questions in the field at this early stage of ML applications. For readers seeking more detailed discussion, \cite{Meher2021} provides a tutorial on the use of deep learning in astronomy, while \cite{Rodriguez2022} presents a systematic scientometric analysis of ML applications in astronomy and astrophysics through text-mining. \cite{Yu2021} offers a comprehensive survey of ML-based light curve analysis for astronomical sources, and \cite{Sen2022} delivers an in-depth review of ML-based processing of astronomical big data. Additionally, \cite{Smith2023} provides a valuable review of artificial intelligence and deep learning in astronomy.

\section{Machine Learning Fundamentals}

ML aims to enable machines to learn from data and make decisions, mimicking human cognitive processes. Unlike traditional programming in astronomy, where explicit physical rules are prescribed, ML identifies patterns, makes decisions, and refines performance using historical data, providing a flexible and adaptive approach. Its tasks include \textit{classification} and \textit{regression}, each crucial in stellar astronomy. Classification identifies and categorizes stellar objects, while regression unravels the physical processes of astronomical phenomena. ML methods are typically divided into supervised and unsupervised learning. Supervised learning uses labeled data to recognize patterns, useful for stellar object identification. Unsupervised learning works with unlabeled data, revealing hidden structures, often used in preprocessing observational data.

\begin{figure*}[htb]
\centering
\includegraphics[width=0.95\textwidth]{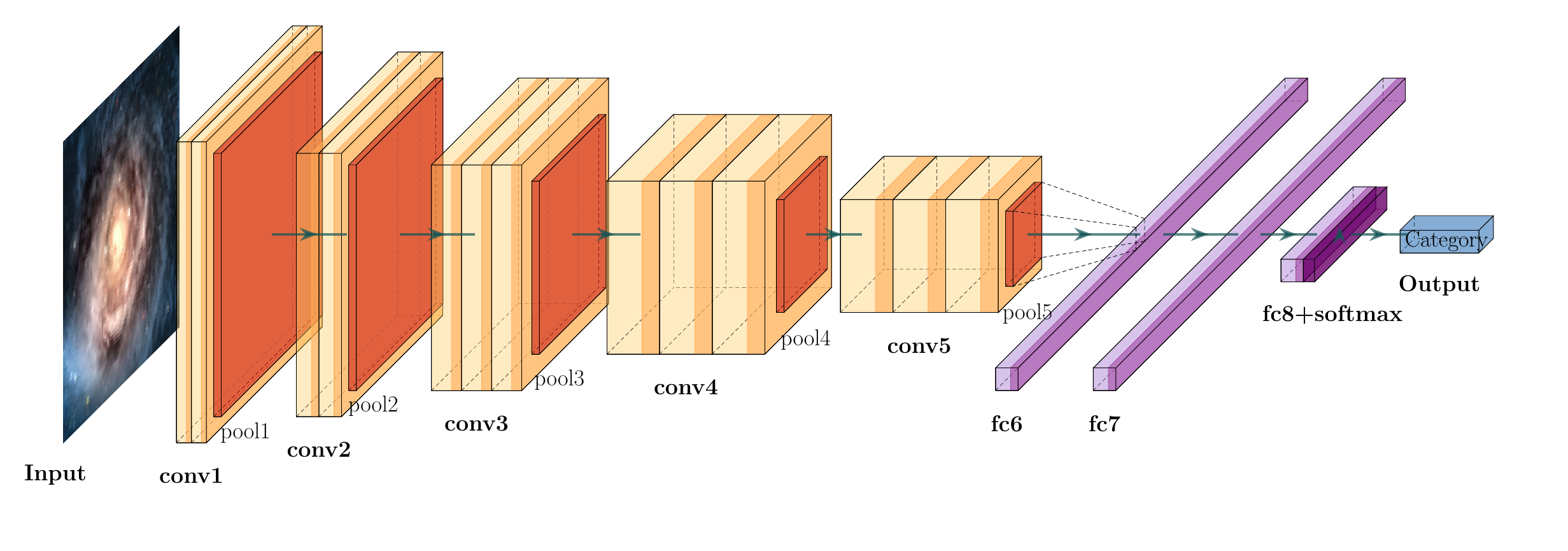}
\caption{Architecture of a CNNs. The diagram shows a typical CNNs pipeline, starting with an input image, followed by multiple convolutional (conv) layers and pooling (pool) operations. The extracted features are progressively refined and passed through fully connected layers (fc6 and fc7) before classification is performed in the final layer using a softmax function.}
\label{F1}
\end{figure*}

Algorithms are foundational for ML. In unsupervised learning, K-means clustering partitions a dataset into K clusters based on similarities, aiding in pattern recognition \citep{Lloyd1982}. For supervised tasks, extreme gradient boosting (XGBoost) is effective for classification and regression, known for its robustness \citep{Chen2016}. Gaussian process regression is a non-parametric approach used for regression, offering insights into uncertainty and variance \citep{Wang2023}. Advanced algorithms frequently used in stellar astronomy are introduced in more details as:

\textbf{Neural networks}: Artificial neural networks (NNs), the backbone of deep learning, consist of interconnected layers of artificial neurons that mimic the brain's neural architecture. These networks process information through input, hidden, and output layers, with learning achieved by adjusting weights and biases via backpropagation to minimize prediction error. Multi-layer perceptrons (MLP) feature multiple layers with feedforward connections, while convolutional neural networks (CNNs) excel in image classification by detecting features from edges to patterns through convolutional and pooling layers. As shown in Figure \ref{F1}, CNNs extract features, reduce spatial dimensions with pooling, and classify images using fully connected layers and a softmax function. This hierarchical approach preserves and refines spatial information, making CNNs ideal for tasks like identifying and classifying stellar objects. Other neural network models include recurrent neural networks (RNNs) for sequential data and probabilistic neural networks (PNNs) for managing uncertainty \citep{Hochreiter1997,Mohebali2020}.

\textbf{Decision Trees and Random Forests}: Decision Trees create a tree-like model by recursively splitting the dataset based on feature values. Nodes represent features, branches indicate decision paths, and leaves correspond to outcomes. Random Forests (RF), an ensemble of decision trees, enhance performance and generalization through voting mechanisms and feature aggregation \citep{Ho1998}. This reduces variance and mitigates overfitting, making it effective for astronomical classification and regression tasks.

\textbf{Support Vector Machines}: Support Vector Machines (SVMs) find an optimal hyperplane that maximally separates classes in a dataset \citep{Huang2011}. SVMs handle non-linear relationships using kernel functions and are suitable for high-dimensional spaces. In stellar astronomy, SVMs classify stars based on observational parameters, discerning complex patterns and handling multi-dimensional data.

\textbf{K-Nearest Neighbors}: K-Nearest Neighbors (KNNs) is an instance-based learning algorithm \citep{Cover1967}. It classifies or predicts data points based on proximity to KNNs. KNNs adapts well to various data distributions and is effective with noisy or less-structured data, making it suitable for star classification and pattern recognition in stellar astronomy. 

\section{Data Sources}
In the realm of ML applications in stellar astronomy, data sources play a crucial role as the foundation for model training and evaluation. The quality and preprocessing of data from these sources are pivotal factors that significantly impact the reliability and effectiveness of a ML model's performance. Stellar astronomers commonly leverage various data sources, each providing rich and diverse datasets essential for training and testing ML models. Here are some primary data sources in stellar astronomy:

\textbf{Sky Surveys}: Sky surveys, such as the SDSS \citep{York2000,Blanton+2017AJ, Abolfathi2018} and the Panoramic Survey Telescope and Rapid Response System (Pan-STARRS \citep{Kaiser2002}), provide extensive catalogs of stars and galaxies, detailing their positions, magnitudes, and spectral characteristics. The European Space Agency's Gaia mission data releases, including Gaia DR1 \citep{Gaia_DR1_2016A&A}, DR2 \citep{Gaia_DR2_2018A&A}, and DR3 \citep{Gaia_DR3_2023A&A}, offer a wide range of astrometric and photometric data for over a billion sources. Gaia DR3 also includes astrophysical parameters and source class probabilities for approximately 470 million and 1.5 billion sources, respectively, encompassing stars, galaxies, and quasars. Additionally, the European Space Agency's Herschel Space Observatory, a facility for far-infrared and submillimetre astronomy \citep{Pilbratt+2010A&A}, has released significant datasets such as the Herschel Astrophysical Terahertz Large Area Survey \citep{Smith+2012MNRAS}, which is the largest key astronomical project containing over 6,000 galaxies. The Herschel infrared Galactic Plane Survey has released high-quality data products (DR1) \citep{Molinari+2016A&A} and DR2 \citep{Elia+2021MNRAS} of massive clumps, which are considered potential progenitors of massive stars.

\textbf{Telescope Observations}: Data collected from ground-based and space-based telescopes, including the Hubble Space Telescope \citep{Kennicutt1998}, and the Visible and Infrared Survey Telescope for Astronomy (VISTA \cite{Emerson2006}), provide detailed observations of stars and their surroundings across various wavelengths.

\textbf{Spectroscopic Surveys}: Surveys like the Large Sky Area Multi-Object Fiber Spectroscopic Telescope (LAMOST \citep{Cui2012}) and the European Southern Observatory's Very Large Telescope \citep{Pasquini2009} offer spectroscopic data crucial for stellar classification and in-depth analysis. The Apache Point Observatory Galactic Evolution Experiment has publicly released infrared spectra of over 650,000 stars, which is a survey of the Milky Way to study the galaxy's chemical and kinematical history \citep{Majewski+2017AJ, Wilson+2019PASP, Abdurrouf+2022ApJS}.

\textbf{Radio Telescopes}: Radio astronomy datasets from instruments like the Atacama Large Millimeter/submillimeter Array\citep{Wootten2009} contribute to the study of stellar radio emissions and various astrophysical phenomena.

\textbf{High-Energy Observatories}: Observatories such as Chandra X-ray Observatory \citep{Weisskopf2002} and Fermi Gamma-ray Space Telescope \citep{Atwood2009} provide valuable data for investigating high-energy astrophysical phenomena associated with stars.

\textbf{Variable Star Observations}: Datasets from the American Association of Variable Star Observers \citep{Smith2012} and similar sources offer information on light curves and other characteristics of stars exhibiting periodic variations, essential for the study of variable stars.

\textbf{Exoplanet Surveys}: Exoplanet Surveys, exemplified by the Transiting Exoplanet Survey Satellite (TESS \citep{Ricker2015}) and Kepler \citep{Borucki2010}, not only contribute significantly to exoplanet discovery but also provide extensive data for understanding stellar properties and behavior.

\section{Identification and classification of stars}
\begin{figure*}[htb]
\centering
\includegraphics[width=0.95\textwidth]{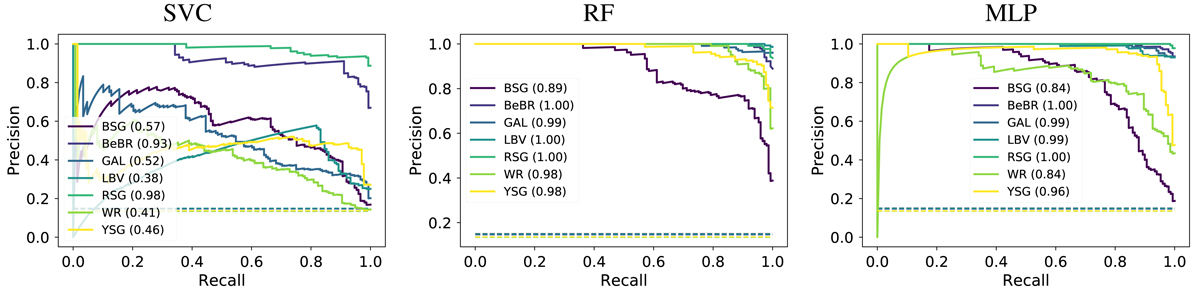}
\caption{This figure shows precision-recall curves for three ML algorithms: SVC, RF, and MLP used to classify different types of massive stars in M31 and M33. Each curve represents the performance of the algorithm for a specific class of stars, with the area under the curve indicating the classification effectiveness\cite{Maravelias2022}.}
\label{F2}
\end{figure*}

We begin by summarizing key works on star identification and classification. In the early 2000s, \cite{Gulati2000} pioneered the use of NNs to predict spectral types and luminosity classes of stars through spectral indices, achieving impressive precision in predicting spectral classes. A more recent study by \cite{Schierscher2011} applied an NNs model to classify stellar spectra from the 7th data release (DR7) of SDSS, demonstrating a high level of agreement with results obtained from other pipelines. \cite{Navarro2012} utilized an NNs for stellar classification based on spectra with relatively low signal-to-noise ratio. For main-sequence stars, \cite{Kuntzer2016} employed NNs to categorize stars into 13 spectral subclasses using single-band imaging based on simulated telescope images, achieving a notably high success rate. In the categorization of massive stars, \cite{Kheirdastan2016} applied PNNs, SVMs, and K-means clustering to classify massive stellar spectra in SDSS data, \cite{Wallenstein2021} found that an SVMs classifier effectively categorizes massive stars into hot, cool, and emission-line stars with high accuracy. \cite{Maravelias2022} introduced a ML-based photometric classifier to identify massive stars in galaxies M31 and M33 . By combining multi-band photometric data and Gaia astrometry, they developed a method to classify different types of massive stars, including blue supergiants(BSG), red supergiants(RSG), and Wolf-Rayet(WR) stars, etc. To evaluate the classifier's performance, they used precision-recall curves shown in Figure \ref{F2}, which demonstrate the effectiveness of three algorithms in distinguishing stellar types.

\begin{figure}[htb]
\centering
\includegraphics[width=0.50\textwidth]{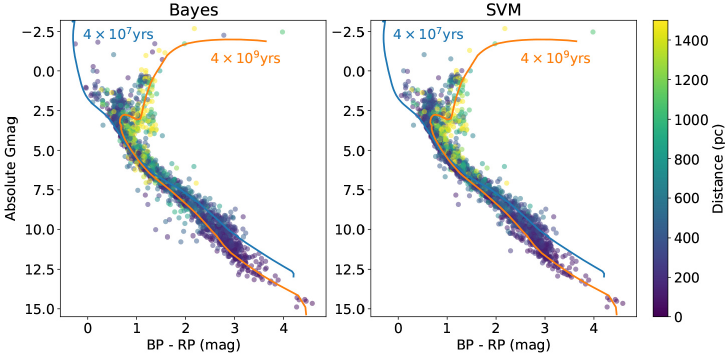}
\caption{Color-magnitude diagram for the identified stellar sources. The color indicates the distance of the sources\cite{Schneider2022}.}
\label{F3}
\end{figure}

\cite{Ostdiek2020} introduced an NNs-driven approach to catalog accreted stars, successfully identifying approximately 767,000 accreted stars within the Gaia DR2 dataset. \cite{Skoda2020} employed CNNs to detect emission-line stars within LAMOST data, achieving high accuracy. \cite{McBride2021} introduced an NNs model to identify pre-main-sequence stars and estimate their ages using data from Gaia DR2 and 2 Micron All-Sky Survey photometry. \cite{Khramtsov2021} used an ensemble of 32 ML models to classify objects in the VISTA EXtension to Auxiliary Surveys data. \cite{Rozanski2022} proposed a CNNs-based technique for the normalization of stellar spectra, yielding results aligned with manual normalization. \cite{Schneider2022} applied ML and Bayesian algorithms to identify stellar counterparts of extended ROentgen Survey with an imaging telescope array X-ray sources. By cross-matching with the Gaia catalog, they successfully classified 2,060 X-ray sources as stellar, achieving near 90\% completeness and reliability. Figure \ref{F3} shows the positions of the classified stellar sources in a color-magnitude diagram.

\section{Inference of Stellar properties}

Besides classification, ML regressive algorithms were used to extract important astrophysical parameters from stellar observations, in order to gain insights into diverse aspects of stellar evolution \citep{Guiglion2024,Ksoll2024}. Early studies, exemplified by \cite{Jones2000}, showcased the accuracy of predicting parameters such as effective temperature $T_\text{eff}$, surface gravity log($g$), and metallicity [M$/$H] using NNs trained on synthetic spectra. \cite{Bellinger2016} introduced a rapid estimation method based on RF for fundamental parameters of main-sequence solar-like stars, leveraging classical and asteroseismic observations. \cite{Hinners2018} applied NNs to predict and classify stellar properties from sparse and noisy time-series data, achieving low-error estimates for stellar density, radius, and $T_\text{eff}$.

\begin{figure}[htb]
\centering
\includegraphics[width=0.5\textwidth]{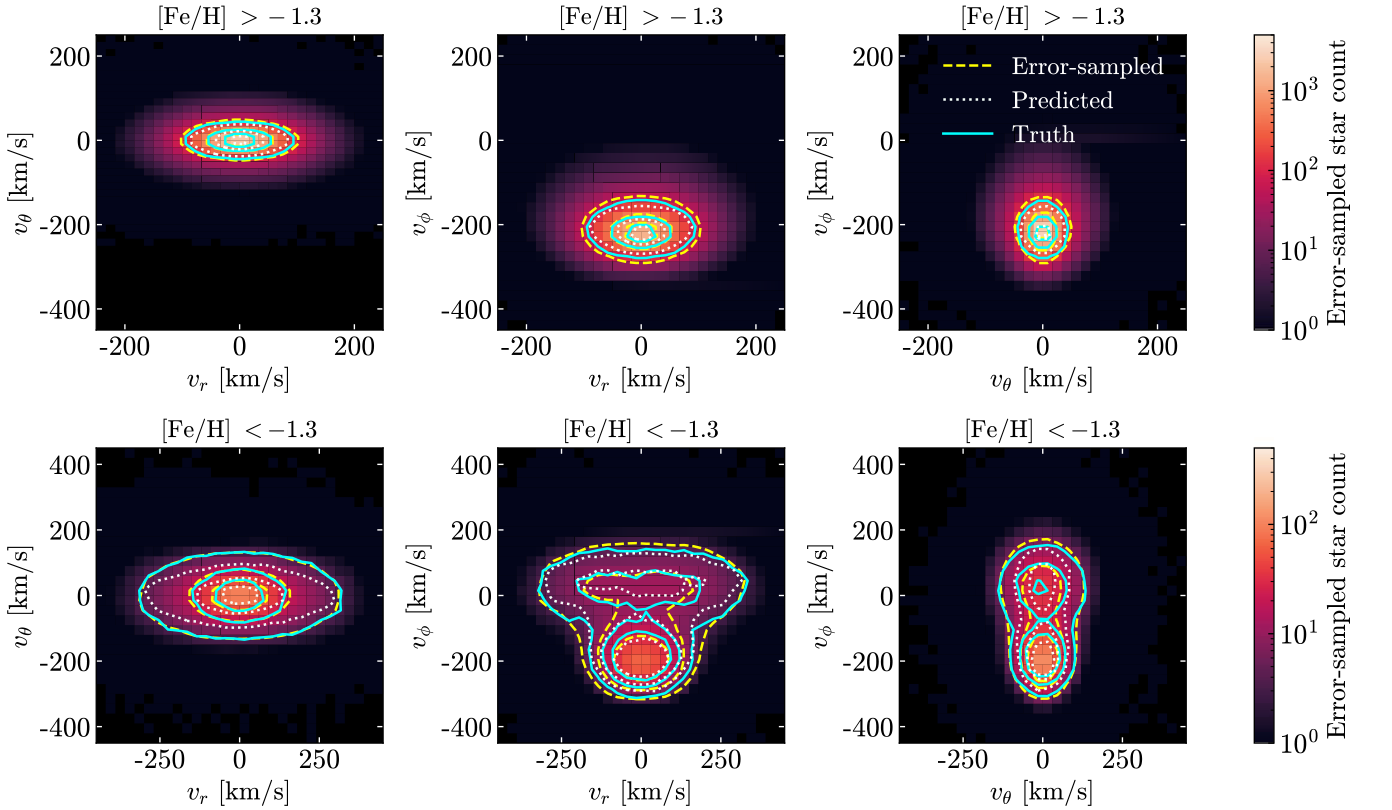}
\caption{2D distributions of the Galactocentric velocity components for stars with different metallicities. The background histogram shows the kinematic distributions predicted by the neural network, with error sampling to reflect uncertainty\cite{Dropulic2021}.}
\label{F4}
\end{figure}

Expanding on these efforts, \cite{Bai2019, Bai2020} employed multiple ML algorithms to predict $T_\text{eff}$ and extinction, achieving precise predictions with minimal errors. \cite{Garraffo2021} presented a deep NNs trained on stellar evolutionary tracks, accurately predicting mass and age from luminosity and $T_\text{eff}$. Advancing the field, \cite{Breton2021,Claytor2022} developed ML pipelines for predicting stellar rotation periods from light curves, showcasing high accuracy. \cite{Dropulic2021} present a NN model to infer the line-of-sight velocities of stars from phase-space coordinates. They trained the network using a simulated Gaia catalog and showed its effectiveness in recovering stellar velocity distributions. Figure \ref{F4} illustrates the performance of their NNs on stellar samples with different metallicities. Their results indicated that the error-sampled distribution more accurately captures the true velocity distribution, especially for metal-poor stars.

\cite{Li2022} derived an ML catalog of stars with mass labels, demonstrating accurate mass predictions. \cite{Qiu2022} developed an NNs model predicting multi-wavelength spectral energy distribution from galaxy parameters and star formation history. \cite{Blancato2022} implemented a CNNs to predict stellar properties from light curves, successfully recovering properties such as surface gravity and rotation period. \cite{Flores2023} present the results of the implementation of a deep learning system capable of estimating $T_\text{eff}$ and log($g$) of O-type stars.

Focusing on the magnetic activity of stars, \cite{Velez2018} employed ML algorithms to measure the mean longitudinal magnetic field in stars from polarized spectra. In the domain of flare detection, \cite{Feinstein2020} developed a CNNs trained to identify flares in TESS short-cadence data, evaluating flare rates as a function of stellar age and spectral type. In a related study, \cite{Vida2021} conducted an experiment to detect flares in space-borne photometric data using NNs, achieving a network capable of distinguishing real flares from typical false signals.

Furthermore,  \cite{Yang2023a} proposed a CNNs-based photometric model to extract stellar fluxes in DECam Legacy Survey DR 9 and SDSS DR 12. \cite{Bonjean2019} proposed an RF-based method to estimate star formation rate and stellar mass in galaxies. Finally, \cite{Li2020} exemplified how ML techniques can explore correlations between stellar parameters beyond prediction, underscoring the versatility and depth of ML applications in stellar astronomy. In the section below, we will conduct a comprehensive survey of studies focused on specific types of stellar objects.

\section {Works on specific types of stellar objects}
In addition to the aforementioned studies addressing the general characteristics of stars, there is a substantial body of research focused on specific types of stellar objects. Below, we provide an introduction to some of the key works in this area, categorized by different types.

\subsection {Binary Stars}
Binary stars (BSs), comprising two stars orbiting a common center of mass, represent the most prevalent form of multiple star systems. ML has become a pivotal tool, supplanting traditional methods in the investigation of BSs \citep{Ding2024}. In the early 2000s, \cite{Weaver2000} pioneered the application of NNs for the two-dimensional spectral classification of BS components. Their study highlighted the efficacy of NNs in classifying a diverse range of stars in binary systems, even accommodating a substantial 3-magnitude difference in brightness, leveraging low-resolution near-infrared spectra.

\cite{Gao2018} employed RF to discern reliable members within the aged open cluster M67. Their classification, rooted in astrometry and photometry data from Gaia-DR2, indicated that over 26.4\% of the main-sequence stars in M67 are binary. \cite{Pirkhedri2012} utilized an NNs to deduce the orbital parameters of spectroscopic BSs, relying on radial velocity data measured from four double-lined spectroscopic binary systems. In the discovery and classification of symbiotic stars, \cite{Akras2019} applied a machine learning (ML) approach to establish robust selection criteria, combining data from the Two Micron All Sky Survey (2MASS) and the Wide-field Infrared Survey Explorer (AllWISE), successfully identifying several new members.

In the realm of light curve categorization for BSs, \cite{Cokina2021} conducted extensive testing with various NNs architectures, revealing that the most effective classifier integrates bidirectional Long Short-Term Memory (LSTM) and one-dimensional CNNs. In a recent advancement, \cite{Zheng2023} introduced a CNNs model designed for identifying double-line spectroscopic binaries. Trained on the cross-correlation function of simulated spectra, the model demonstrated an impressive accuracy.

\subsection {Supernovae}
Supernovae, captivating stellar explosions marking the end of massive stars or white dwarf nuclear fusion runaways, present challenges in identification and classification due to their transient nature, voluminous data, and redshift effects. ML offers solutions to these challenges \citep{Yin2024}. For instance,  \cite{Lochner2016} proposed a multifaceted ML pipeline for supernova photometric classification, integrating algorithms like naive Bayes, KNNs, SVMs, neural networks, and boosted decision trees.  \cite{Charnock2017} used RNNs for supernova classification, emphasizing ML's ability to learn from light curves with sensitivity to training data.

\cite{Fremling2021} introduced an RNNs-based method for classifying low-resolution thermonuclear supernovae (SNe Ia), achieving a low false-positive rate. \cite{Ishida2017} showcased MLs's proficiency in identifying SNe Ia subtypes, and \cite{Qu2021} presented a CNNs model for Ia supernovae classification without redshift information. \cite{Qu2023} presented a CNNs-based appraoch for predicting full redshift probability distributions from multi-band type Ia supernova lightcurves, based on both simulated data and observed data from SDSS and Vera C. Rubin Legacy Survey of Space and Time. They showed improvements over predictions from traditional methods. \cite{Muthukrishna2019} developed CNNs software outperforming in type, age, redshift, and host galaxy classification. \cite{Gomez2020} used RFs to identify hydrogen-poor superluminous supernovae, achieving a maximum purity of 85\%.

\cite{Hsu2022} performed photometric classification of Type I superluminous supernovae, and \cite{Morgan2022} designed deep neural networks for swift lensed supernovae (LSNe) identification. ML contributes to understanding supernova physical properties. \cite{Ansari2022} used an NNs to infer dust properties in supernovae. \cite{Harada2022} trained NNs for Eddington tensor prediction. \cite{Vogl2020} presented an ML emulator for Type II supernovae physical properties. \cite{Hu2022} used LSTM NNs for Type Ia supernovae spectral time series analysis. \cite{Huber2022} measured LSNe Ia time delays with ML models.

Regarding astrophysical properties of supernovae, \cite{Chao2022} developed CNNs models for core-collapse supernova properties. \cite{Tsang2022} used RF for core-collapse explosion outcome prediction. \cite{Karpov2022} developed a physics-informed CNNs for accurate turbulent pressure prediction in supernovae. \cite{Kerzendorf2021} presented an NNs model for supernova spectrum synthesis, offering accuracy with significant speedup.

\subsection{Dwarfs}
\begin{figure}[htb]
\centering
\includegraphics[width=0.50\textwidth]{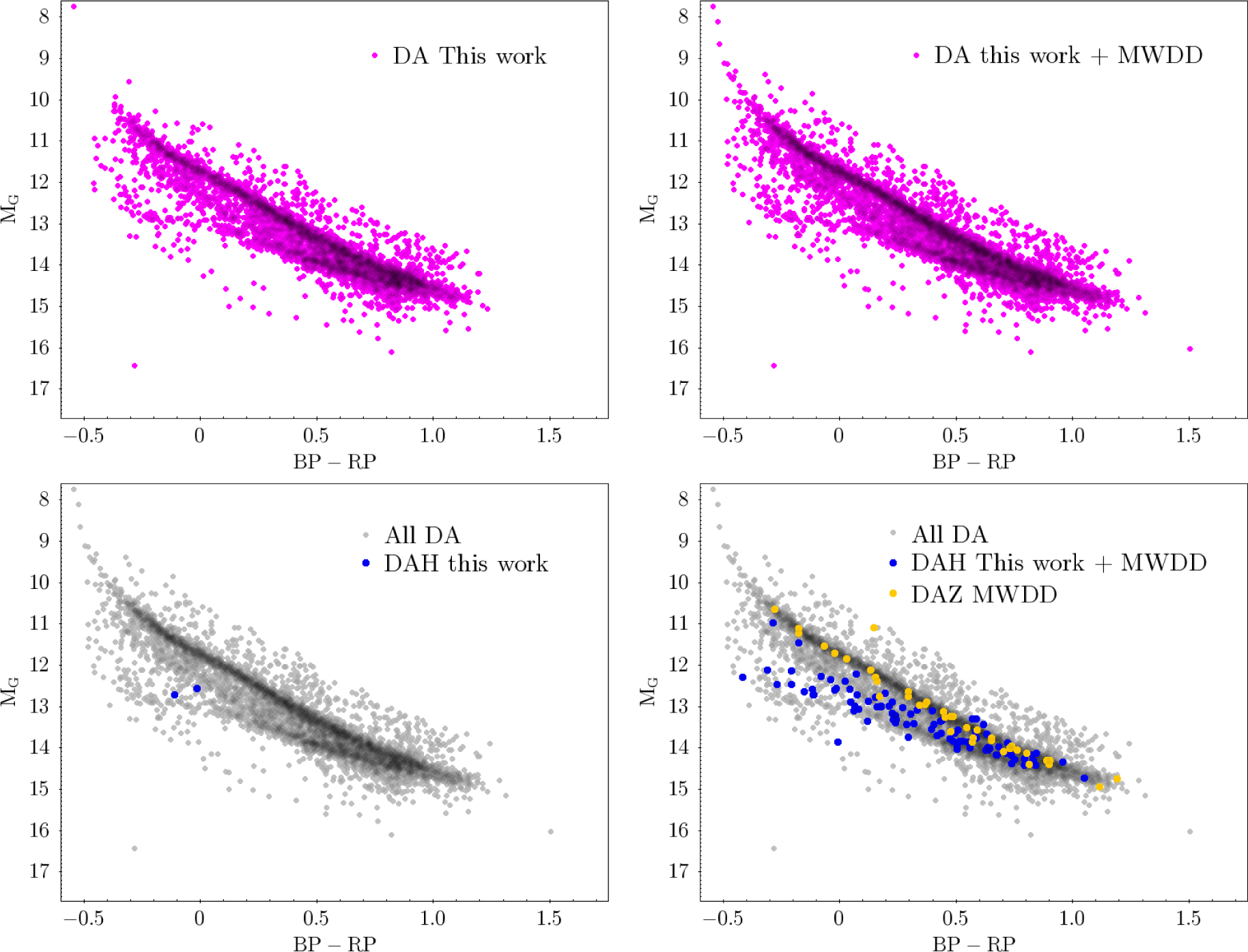}
\caption{Distribution of DA white dwarfs classified in this study (left panels) and the entire population (right panels) in the Gaia diagrams\cite{Garcia-Zamora2023}.}
\label{F5}
\end{figure}

A dwarf star, characterized by its relatively small size and low luminosity, encompasses various types, including main-sequence stars such as yellow, orange, and red dwarfs, as well as brown, white, and other dwarf stars. Notably, studies over several decades have explored diverse ML applications in stellar classification \citep{Tan2023}. \cite{Garcia-Zamora2023} proposed a RF-based model for classifying white dwarfs from Gaia spectra. By utilizing Hermite coefficients as input features, they successfully assigned spectral types to 9,446 previously unclassified white dwarfs, significantly increasing the number of white dwarfs of various types within 100 parsecs. Figure \ref{F5} illustrates the distribution of the white dwarfs classified in this study within the Gaia HR diagrams.

In 1997, \cite{Gulati1997} utilized NNs to deduce observational properties, specifically effective temperatures, of G-K dwarfs. Their approach involved analyzing a library of synthetic spectra. \cite{Karnavas2020} contributed to the field by developing an ML tool using linear regression algorithms from the "scikit-learn" package. This tool successfully predicted stellar parameters, including effective temperature and Fe/H ratio, for M dwarf stars.

\cite{Torres1998} employed an NNs for the classification of spectroscopically identified white dwarfs. In a related context, \cite{Feeney2005} utilized a set of 440 NNs to identify fast classical novae, occurring in binary systems with a short-period orbit between a white dwarf and a main sequence star. \cite{Yang2020} applied an NNs to estimate $T_\text{eff}$ and log($g$) of white dwarfs. The model was directly trained and tested on the continuum-normalized spectral pixels of DAs spectroscopically confirmed in SDSS DR7, DR10, and DR12. The results demonstrated accuracy without relying on complex theoretical frameworks. \cite{Tan2023} presented a deep transfer learning model for the  identification of white dwarfs from spectra in LAMOST DR9. Their method has identified 6317 candidates, among which 5014 were cross-validated with known white dwarfs.

\cite{Buitrago2022} introduced a ML methodology incorporating principal component analysis and SVMs algorithms. This approach identified 7,827 new candidate ultracool dwarfs from the Javalambre Photometric Local Universe Survey (J-PLUS) second data release. \cite{Oreshenko2020} integrated a RF algorithm into classical models for spectral analysis of brown dwarfs, enhancing the efficiency of the analysis. \cite{Avdeeva2023} used ML algorithms such as RF, XGBoost, SVMs and TabNet on PanStarrs DR1, 2MASS and WISE data for the search of L \& T brown dwarfs. In a work of \cite{Lueber2023}, the RF method enables analyzing the predictive power of 14 model grids of brown dwarfs.

Hot subdwarf stars, crucial for studying binary evolution and atmospheric diffusion processes, have been a focus of ML applications. \cite{Bu2019} proposed a method using CNNs and SVMs to select hot subdwarf stars from LAMOST spectra, achieving an F1 score of 76.98\%. Similarly, \cite{Tan2022} introduced a CNNs-based method for identifying hot subdwarf stars in the spectral data of LAMOST DR7, resulting in a catalog of 2,393 hot subdwarf candidates, of which 2,067 have been confirmed. These endeavors align with an early work by \cite{winter2004}.

\subsection{Young Stellar Objects}
ML actively contributes to the exploration of Young Stellar Objects (YSOs), which are stars in the early stages of evolution encompassing protostars and pre-main-sequence stars. Notable studies demonstrate the efficacy of various ML algorithms in this domain. For instance, \cite{Miettinen2018} employed eight distinct ML algorithms to classify YSOs based on evolutionary stages in Orion. The study revealed that the Gradient Boosting Machine (GBM) exhibited superior performance, achieving an accuracy of 82\% in comparison to classes determined through spectral energy distribution analysis.

\cite{Melton2020} utilized a RF model to identify YSOs in the Lupus star-forming region, leveraging data from Gaia and Spitzer surveys. The model successfully identified 27 Lupus candidates, demonstrating reliability for subsequent spectroscopic follow-up. \cite{Chiu2021} developed a NNs for classifying stars, galaxies, and YSOs based on infrared spectra. Trained on data labeled in the molecular cores to planet-forming disks catalog and applied to Spitzer Enhanced Imaging Products by a spectrum classifier of astronomical objects, the model exhibited high precision and recall for YSOs.

\cite{Cornu2021} presented an NNs model for YSOs identification from the Spitzer Infrared (IR) survey. The model achieved a recovery rate above 90\% and 97\% for CI and CII YSOs, respectively, with precision exceeding 80\% and 90\% for their most general results. For regression tasks, \cite{Olney2020} trained a CNNs model using labels of YSOs and low-mass main-sequence stars generated through photometry. The model was employed to predict log($g$), $T_\text{eff}$, and Fe/H for stars with spectra in the SDSS DR14 dataset, demonstrating its versatility in handling diverse aspects of YSOs research.

\subsection{Variable stars}
The study of variable stars, those that undergo changes in brightness over time, is a crucial aspect of stellar astronomy, providing insights into stellar properties, evolutionary stages, and system dynamics. ML techniques have played a significant role in advancing research in this field \citep{Qiao2024}. For instance, \cite{Richards2011} pioneered the application of the RF algorithm for the classification of variable stars, particularly those with sparse and noisy time-series data. Testing their model on a 25-class dataset of 1,542 well-studied variable stars, they achieved a commendable 22.8\% error rate. Building on this, \cite{Szklenar2022} employed a multiple-input NNs trained on Optical Gravitational Lensing Experiment-III data, achieving accuracies ranging from 89\% to 99\% for most main classes of variable stars.

\cite{Vilalta2013} presented an ML model for classifying two subtypes of Cepheid variable stars, utilizing data alignment and maximum likelihood. Their method, demonstrated on two galaxy datasets, showcased its effectiveness. \cite{Tsang2019} introduced a periodic light curve classifier combining a RNNs autoencoder for unsupervised feature extraction and a dual-purpose estimation network for supervised classification and novelty detection. The model, incorporating photometric features, exhibited high precision (0.90), recall (0.96), and an F1 score of 0.93. \cite{Rodriguez2023} developed a supervised ML algorithm to automate the classification of T Tauri stars in TESS light curves.

For RR Lyrae stars (RRLs), \cite{Cabral2020} developed an ML-based procedure for their identification in the VISTA Variables in the Via Lactea (VVV) Survey, achieving an average recall of 0.48 and average precision of 0.86. \cite{Dekany2020} utilized a deep near-IR variability search with a RNNs classifier based on VVV survey data to identify fundamental-mode RRLs, achieving around 99\% precision and recall for high signal-to-noise ratio light curves. Additionally, their research group introduced a novel method for metallicity estimation of fundamental-mode RRs using deep learning, obtaining low mean absolute error and high regression performance.

\cite{Sesar2017,Stringer2019} employed ML-based template fitting techniques for RRLs identification in the PanSTARRS1 and Dark Energy Survey datasets, respectively, obtaining accurate period estimates and classification results. \cite{Zhang2020} applied cost-sensitive RF for imbalanced learning to preselect RRLs candidates from SDSS data, identifying 11,041 photometric candidates with spectral type A and F. \cite{Mombarg2021} trained NNs to predict theoretical pulsation periods of high-order gravity modes and to predict the luminosity, effective temperature, and surface gravity for a given mass, age, overshooting parameter, diffusive envelope mixing, metallicity, and near-core rotation frequency of Gamma Doradus stars. The NNs approach developed allows the derivation of stellar properties dominant for stellar evolution.

\subsection{Metal-poor stars}
The observation of metal-poor (MP) stars offers a unique glimpse into the early cosmic evolution. ML techniques have proven invaluable in enhancing the precision of estimating crucial stellar parameters for MP stars, including effective temperature, metallicity ([Fe/H]), and log($g$). In a pioneering effort, \cite{Prieto2012} utilized an NNs model to estimate [Fe/H] and intrinsic (B-V) colors of MP stars identified in a survey conducted with the Isaac Newton Telescope and the Intermediate Dispersion Spectrograph. Despite a comparatively lower effective yield of MP stars, their approach demonstrated a high rate of discovery per unit of telescope time. 

\cite{Snider2001} extended the application of NNs to estimate $T_\text{eff}$, log($g$), and [Fe/H] for Galactic F- and G-type stars. Their model, even without considering photometric information, achieved competitive accuracies compared to results obtained through fine analysis of high-resolution spectra. \cite{Giridhar2013} further refined the use of NNs by training on a library of 167 medium-resolution stellar spectra, achieving an impressive accuracy of 0.3 dex in [Fe/H], 200 K in temperature, and 0.3 dex in log($g$). \cite{Whitten2019} developed an NNs-based photometry pipeline for estimating $T_\text{eff}$ and [Fe/H] in a subset of stars within the J-PLUS footprint. They validated its utility as a standalone tool for photometric estimation and pre-selection of MP stars.

\cite{Xie2021} proposed a CNNs model for determining atmospheric parameters of very MP stars, outperforming other widely used methods when tested on LAMOST spectra. \cite{Hughes2022} employed ML algorithms to identify 54 candidates of extremely MP (EMP) stars from a vast dataset of high-resolution stellar spectra in the GALactic Archaeology with Herschel Multi-tiered Extragalactic survey. Their findings align with previous work, validating the consistency of the magnitude-limited metallicity distribution function in their sample. \cite{Hartwig2023} introduced a novel ML method utilizing SVMs to classify EMP stars into mono- or multi-enriched categories. They identified Fe, Mg, Ca, and C as the most informative elements for this classification.

\subsection{Chemical peculiars and others}
\textbf{Chemical peculiars}: \cite{Shang2022} undertook a comprehensive investigation into class-one and class-two chemical peculiars (CP1 and CP2) within the 8th data release (DR8) of LAMOST. Their methodology involved the utilization of the XGBoost algorithm, leading to the successful identification of a substantial total of 6,917 CP1 and 1,652 CP2 stars. In a related study, \cite{Guo2023} harnessed ML algorithms, including the Light GBM and SVMs, to construct precise and efficient classifiers for barium stars. Utilizing low-resolution spectra from LAMOST, their approach demonstrated effectiveness in distinguishing barium stars. Additionally, \cite{Hartogh2023} developed ML algorithms that utilized the abundance pattern of Ba stars as input to classify the initial mass and metallicity of the companion stars associated with barium stars. This approach, based on stellar model predictions, successfully identified the key properties for 166 out of the 169 barium stars within the stellar sample.

\textbf{Be and B[e] Stars}: \cite{Skoda2016} utilized a Spark-based semi-supervised ML approach to automate the identification of Be and B[e] stars in LAMOST spectra. The resulting list of candidates not only included numerous Be stars but also presented a diverse array of objects resembling spectra from quasars and blazars, along with various instrumental artifacts. \cite{Ortiz2017} trained a RF classifier to identify 50 Be star candidates in the OGLE-IV Gaia south ecliptic pole field, with four of them exhibiting infrared colors consistent with Herbig Ae/Be stars. \cite{Vioque2020} applied ML techniques to approximately 4 million sources from Gaia DR2, 2MASS, WISE, and IPHAS or VPHAS+. This resulted in a catalogue of 8,470 new pre-main sequence candidates and another catalogue of 693 new classical Be candidates, among which at least 1,361 sources are potentially new Herbig Ae/Be candidates.

\textbf{Carbon stars}: \cite{Li2018} harnessed the power of a supervised rank-based Positive-Unlabeled learning algorithm to effectively pinpoint 2,651 carbon stars within a dataset exceeding 7 million spectra from the 4th Data Release (DR4) of LAMOST. These stars underwent further classification into five distinct subtypes based on a unique set of spectral features.

\textbf{Primordial Stars}: \cite{Wells2021} introduced a CNNs-based model tailored for predicting localized primordial star formation. The model's outcomes align seamlessly with intricately detailed cosmological simulations.

\textbf{Giants}: \cite{Hon2018} used NNs to detect oscillations in red giants with high accuracy. \cite{Yuan2023} proposed a CNNs-based method for the search of red giants by enhancing feature extraction from images from the Sky Mapper Southern Survey and the LAMOST. The results show that the ML method outperforms traditional approaches across most performance metrics \citep{Dhanpal2023}.

\subsection{Related ISM objects}
Moreover, a number of studies have been conducted on related ISM  objects. Below, we introduce some of these works, categorized as follows:

\textbf{Protoplanetary Disks}: \cite{Auddy2022} introduced a Bayesian NNs explicitly designed to forecast planet mass by analyzing observed gaps in protoplanetary disks. Trained on a dataset derived from disk-planet simulations, this model exhibits a notable ability to discriminate between uncertainties linked to the deep-learning architecture and those inherent to input data due to measurement noise. Additionally, \cite{Smirnov2022} trained a KNNs regressor to predict the chemistry of protoplanetary disks based on specific physical conditions, providing a practical approach for Bayesian fitting of line emission data to extract disk properties from observations. \cite{Telkamp2022} employed a ML framework to predict images of edge-on protoplanetary disks, demonstrating that this framework generates synthetic images 2-3 orders of magnitude faster than traditional radiative transfer calculations. Furthermore, \cite{Kaeufer2023} trained two NNs to predict spectral energy distributions of protoplanetary disks. The results indicate that the NNs can accurately predict properties of protoplanetary disks with significantly higher efficiency than conventional methods. 

\begin{figure}[htb]
\centering
\includegraphics[width=0.5\textwidth]{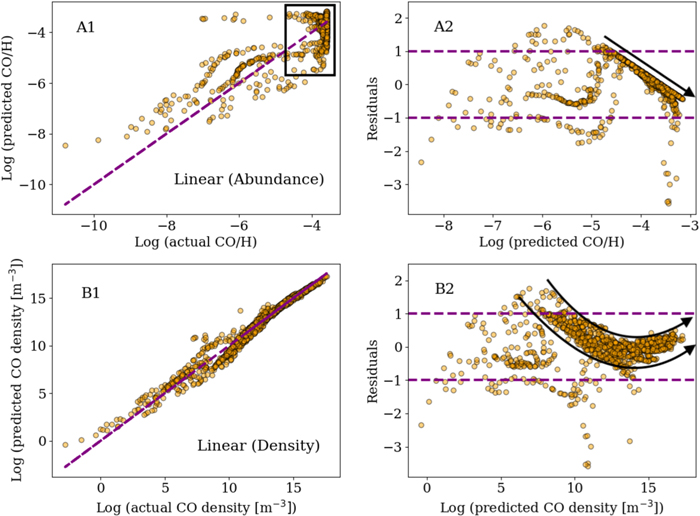}
\caption{(A1) Comparison of predicted CO abundances (ML model) and actual CO abundances. (A2) Residuals plotted against predicted CO abundances, defined as the difference between log(actual CO abundance) and log(predicted CO abundance). (B1-2) The same analysis performed using CO densities instead of abundances\cite{Diop2024}.}
\label{F6}
\end{figure}

\cite{Diop2024} proposed a ML-based approach to study the spatial density distribution of CO in protoplanetary disks. They employed a neural network regression model, incorporating parameters such as gas temperature, gas density, cosmic ray ionization rate, X-ray ionization rate, and ultraviolet flux, to predict the spatial density of CO and quantify the associated uncertainties. Figure \ref{F6} illustrates the relationship between predicted values, actual values, and residuals after applying multiple linear regression to the test set. By analyzing the linear and nonlinear trends, it becomes apparent that the model performs less effectively at higher CO abundances and exhibits clustering phenomena in the residuals.

\textbf{Planetary Nebulae}: \cite{Abans1996} employed hierarchical cluster analysis and an NNs to classify Planetary Nebulae(PN). The outcomes of their research demonstrated that the NNs is a reliable tool for the classification of PN. In a subsequent study by \cite{Akras2019a}, 2MASS and AllWISE photometric data were combined with a classification tree algorithm to establish infrared criteria robust enough to distinguish PNe from other classes of H $\alpha$ emitters. By applying these criteria to the IPHAS list of PN candidates, as well as the entire IPHAS and VPHAS+ DR2 catalogs, they identified 141 sources. Among them, 92 were known PNe, 39 were new, and 10 were classified as H ii regions, Wolf-Rayet stars, AeBe stars, or YSOs.

\textbf{Cold Neutral Medium}: \cite{Murray2020} unveiled a CNNs model engineered to extract HI properties across expansive regions without relying on optical depth information. The findings underscore the CNNs's precision in predicting the fraction of cold neutral medium and optical depth, validated through comparisons with direct constraints derived from 21 cm absorption.

\textbf{Feedback Bubbles}: In the search of Feedback Bubbles, \cite{Beaumont2014} introduced Brut, a RF model designed to detect bubbles in infrared images sourced from the Spitzer Space Telescope. Trained on bubbles identified by citizen scientists from the Milky Way Project, Brut exhibited proficiency comparable to expert astronomers in bubble identification. \cite{Xu2017} enhanced Brut's performance in identifying bubbles from synthetic dust observations by retraining it with simulated bubbles, and subsequently implemented a deep NNs to recognize wind-driven shells and bubbles, utilizing data from magnetohydrodynamic simulations. In a subsequent work, this method was extended to identify stellar feedback bubbles in CO emission line spectra \citep{Xu2020}.

\textbf{ISM models}: \cite{Palud2023} proposed a NNs-based approach to derive approximations of the model for inferring physical conditions of ISM from the observations of atomic and molecular tracers. Their proposed strategies produce networks that outperform traditional interpolation methods in terms of accuracy by a factor of 2.

\textbf{Molecular clouds}: \cite{Luo2024} introduced a semi-supervised CNNs model for verifying molecular clumps. The model was trained on a dataset with three different density regions, achieving high accuracy, recall, precision, and an F1 score on the test set, which closely match the results from manual verification. \cite{feng2024} propose a method based on Gaussian decomposition and graph theory, to identify molecular cloud structures and address the over-linking issue. Their method effectively distinguishes gas structures in dense regions, preserves most of the flux without requiring global data clipping or assumptions about structural geometry, and can accommodate multiple Gaussian components for complex line profiles.

\section{Chronological analysis}

\begin{figure}[htb]
\centering
\includegraphics[width=0.50\textwidth]{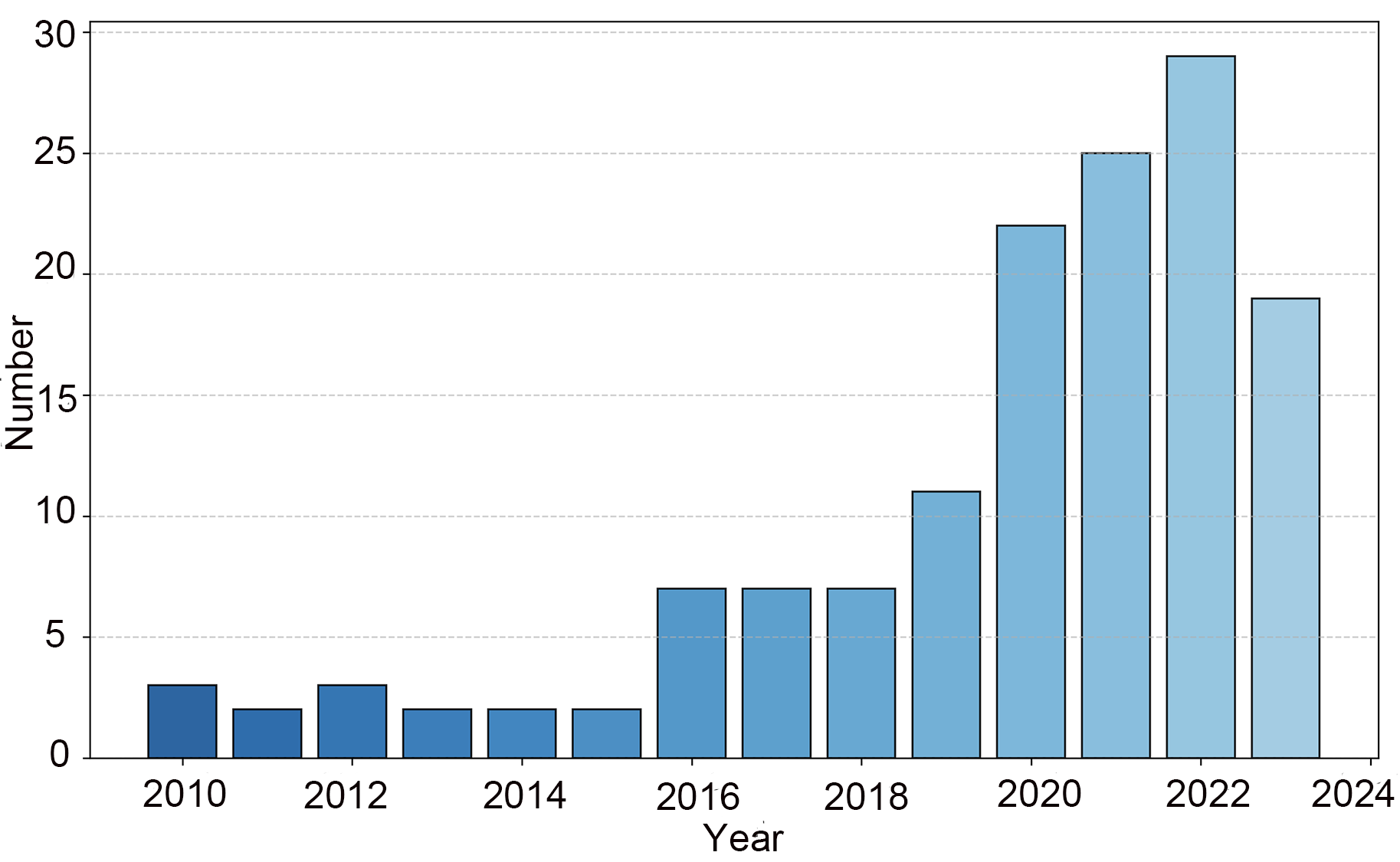}
\caption{Evolution of the number of publications on ML in stellar astronomy from 2010 to 2024.}
\label{F7}
\end{figure}

A scientometric analysis utilizing text mining techniques was conducted on the aforementioned papers on the applications of ML in stellar astronomy. VOSviewer software and data from the Web of Science were employed to elucidate the evolution of publications in this research area, their distribution by location (including co-authorship), and the most relevant topics addressed. As shown in Figure \ref{F7}, from 2010 to 2018, the number of relevant publications was relatively low and showed slow growth. However, there has been a significant increase in the quantity of literature since 2019. Notably, peaks were observed in 2021 and 2022, with over 20 and 25 publications per year, respectively.

\begin{figure}[htb]
    \centering
    \includegraphics[width=0.5\textwidth]{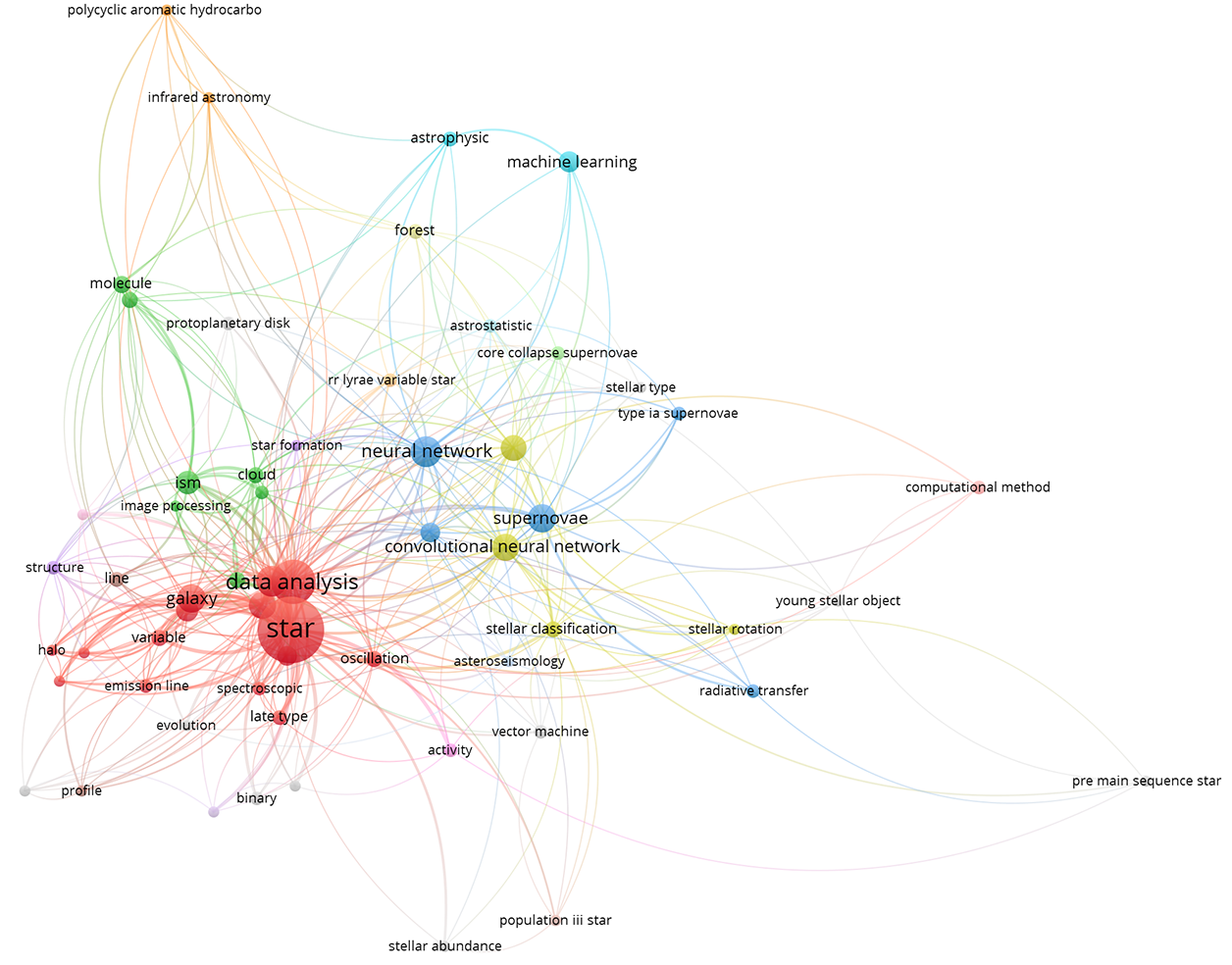}
    \caption{Map of topics (co-occurrence of article keywords).}
    \label{F8}
\end{figure}

VOSviewer was then utilized to conduct a topic analysis through semantic analysis, creating a map that illustrates the main subjects of interest related to the application of ML in astronomy and astrophysics, and their interrelationships. Initially, 281 relevant keywords were identified in the Web of Science database; after setting a minimum threshold of three keyword occurrences, only 59 remained. These were subsequently classified into five main clusters based on their similarities, with each cluster represented by a different color, as shown in Figure \ref{F8}. The clustering process grouped terms that frequently co-occur, reflecting their interconnection. The connections between nodes highlight the interdisciplinary nature of modern astrophysical research, where computational methods and data analysis techniques are essential for understanding stellar and galactic phenomena. The size and color intensity of each node correlate with the frequency and centrality of the term within the research literature, indicating its significance in the field.

\section{Conclusions and perspectives}

As summarized in this review, ML has significantly contributed to advancing the study of star formation and the ISM evolution. ML techniques, including unsupervised learning and deep NNs, are increasingly being applied to large-scale observational datasets, enabling the identification of stars, molecular clouds, and their underlying physical properties. These methods are becoming essential for identifying the initial conditions conducive to star formation, modeling the complex feedback mechanisms of newly formed stars, and simulating the turbulent and magnetic processes that affect cloud collapse and star formation efficiency. Additionally, ML enhances the efficiency of star formation and galactic evolution simulations by improving computational models and enabling more accurate predictions of stellar populations and ISM dynamics.

Nevertheless, it is still premature to fully recognize the transformative potential of ML in addressing fundamental questions in stellar astronomy, particularly those related to the formation, life cycles, and chemical contributions of stars to galactic evolution. Several challenges continue to hinder the full application of ML in this domain. Data quality remains a critical issue, as many ML models are trained on noisy or incomplete datasets, which undermines the accuracy of their predictions. Furthermore, the ``black box'' nature of many ML models, particularly deep learning approaches, complicates the interpretability of results, making it difficult to extract the underlying physical principles driving the predictions. These challenges underscore the need for improved data preprocessing techniques and the development of more interpretable models that are consistent with astrophysical theories.

\newpage
\bibliographystyle{plainnat}

\end{document}